\pdfoutput=1

\documentclass[10pt,letterpaper]{article}
\usepackage{cite}
\usepackage{graphicx}
\usepackage[caption=false,font=footnotesize]{subfig}

\begin{document}

\title{Quantum Metropolitan Optical Network based on Wavelength Division Multiplexing}

\author{{\small A. Ciurana\footnote{Research Group on Quantum Information and Computation, Universidad Polit\'{e}cnica de Madrid, Spain}, J. Mart\'{i}nez-Mateo$^*$, M. Peev\footnote{Optical Quantum Technology, Austrian Institute of Technology, Austria.}, A. Poppe$^\dagger$, N. Walenta\footnote{Group of Applied Physics, Universit\'{e} de Gen\'{e}ve, Switzerland.},}\\ {\small H. Zbinden$^\ddagger$, and V. Mart\'{i}n$^{*,}\footnote{vicente@fi.upm.es}$}}

\date{}

\maketitle

\begin{abstract}
Quantum Key Distribution (QKD) is maturing quickly. However, the current approaches to its application in optical networks make it an expensive technology. QKD networks deployed to date are designed as a collection of point-to-point, dedicated QKD links where non-neighboring nodes communicate using the trusted repeater paradigm. We propose a novel optical network model in which QKD systems share the communication infrastructure by wavelength multiplexing their quantum and classical signals. The routing is done using optical components within a metropolitan area which allows for a dynamically any-to-any communication scheme. Moreover, it resembles a commercial telecom network, takes advantage of existing infrastructure and utilizes commercial components, allowing for an easy, cost-effective and reliable deployment. 
\end{abstract}


\section{Introduction}
\label{sec:intro}

Quantum key distribution allows two distant parties to grow a secret key: an initial shared secret key can be made arbitrarily large while avoiding any information leakage. This is a information theoretic secure scheme based on the laws of quantum mechanics. The price to pay for such a high level of security is the usage of a symmetric key protocol with point-to-point connections \cite{Gisin_02}. Both parties have to be connected through a quantum and a classical but authenticated channel, typically implemented by dedicated optical fiber links. The technology is mature enough for commercialization \cite{idQuantique, Toshiba, Magiq, Sequrenet, AIT}, and long-term practical settings have already been tested \cite{SwissQuantum, Mirza_10, Jouguet_12a}. However, taking this concept to a network setup results in the need to use a completely separated optical infrastructure for QKD \cite{Elliot_02, Peev_09, Stucki_11, Sasaki_11, Kitayama_11} which considerably increases its cost. Sharing an already deployed network and as much commercial technology as possible is then a must for the widespread adoption of QKD as a mainstream security technology.

Nowadays, most telecom networks have adopted the optical paradigm \cite{Chen_99}. The use of passive optical technology is attracting interest in these networks since the absence of active components in the optical pathway, such as amplifiers or electro-optical converters, allows for a more robust and reliable network \cite{Lee_06}---albeit at the cost of some flexibility. From the quantum perspective this means that a unique, uninterrupted optical path can be set between two users and then used as quantum channel, i.e. quantum states can be transmitted in the network without being disrupted. Therefore, it opens the way for integrating QKD systems in commercial telecom networks; this has been a recurring issue in the last years \cite{Townsend_94, Townsend_97b, Townsend_98, Kumavor_05, Fernandez_07, Maeda_09, Lancho_09, Choi_10, Capmany_10, Choi_11, Razavi_12}. It should be further mentioned that the discussed technology is mainly found in networks up to a metropolitan area scale (e.g. access networks and metro backbones), which in turn are the perfect market for QKD: they serve final users and the losses are compatible with the budget and key rate of actual QKD systems \cite{Dixon_08, Takesue_07, Stucki_09b, Wang_12, Namekata_11, Jouguet_12b, Treiber_09}.

Furthermore, wavelength division multiplexing (WDM) \cite{CWDM_03, DWDM_12} is becoming a dominant technology in standard telecom networks. This allows to share efficiently a common optical infrastructure among multiple users \cite{Brackett_90}. Ideally, a QKD system could communicate in these networks using a dedicated wavelength (i.e. a channel) for its quantum signal. Unfortunately, the transmission of single-photon pulses in a fiber together with strong, classical signals (carrying $\approx 10^7$ photons per pulse) is disturbed by the noise generated by the latter. The coexistence of quantum and classical channels is thus limited to just a few of them \cite{Townsend_97a, Xia_06, Rohde_08, Chapuran_09, Lancho_09, Qi_10, Peters_09, Eraerds_10, Patel_12}, especially when they operate in the same spectrum band.

The objective of this work is to devise a technologically realistic and cost-effective QKD network, able to overcome the major roadblocks in the way towards a broader acceptance of QKD technology. The network design is inspired by the technologies and topologies of commercial telecom networks in order to use existing deployed infrastructures (e.g. dark fibers) and commercial components, such that the deployment and running costs are as low as possible and remain competitive with other high security network services.  To this end, QKD devices are wavelength multiplexed in order to share resources. This includes quantum and classical signals, the latter being either generated for the stabilization of the quantum channel or for other QKD purposes like key distillation or encryption. Communications between QKD devices are routed using passive optical components in contrast to trusted repeaters \cite{Peev_09}. However, this fully passive version only works with static QKD links. In the case that an any-to-any scheme is required, optical switches must be added for dynamic routing. Finally, a network prototype based on the proposed model has been designed and deployed for testing purposes. The present approach focuses on prepare-and-measure QKD schemes that fall into two main classes according to the standard classification \cite{Scarani_09}: discrete variables QKD and distributed phase-reference pulse QKD. Its extension to other QKD schemes such as continuous variable QKD and entangled photon-pairs QKD might be possible but lies beyond the scope of the present work.

The paper is organized as follows. Sec.~\ref{sec:man} reviews the architecture and principle of operation of modern metropolitan optical networks. In Sec.~\ref{sec:QKD-MON} we discuss the proposed multiplexing scheme and the modifications required on the network nodes in order to use quantum signals. A prototype of a metropolitan QKD network is described and characterized in Sec.~\ref{sec:prototype}. Finally, we summarize the discussion and outline some future improvements in Sec.~\ref{sec:conclusions}.

\section{Metropolitan optical network} 
\label{sec:man}

\begin{figure}[htbp]
\centering
\includegraphics[width=0.7\textwidth]{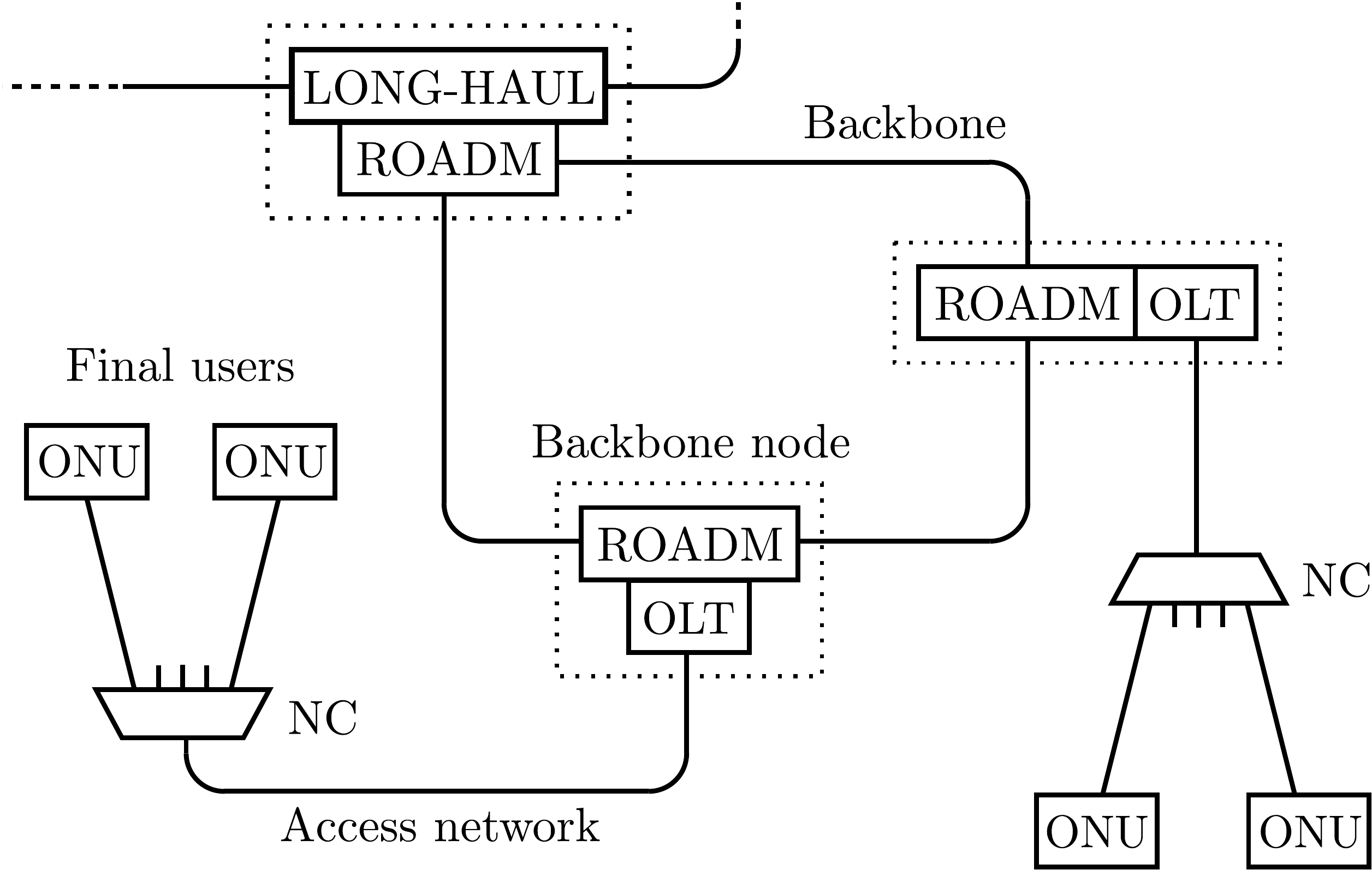}
\caption{Typical architecture and topology used in the \textit{canonical} metropolitan optical network considered in this work. A core network, the backbone, with the highest capacity links set up in a ring, is connected with the final users through one-to-many passive access networks. The network component (NC, typically an splitter or multiplexing device) is usually located near the users (optical network units or ONUs) in order to minimize the amount of non-shared fiber used. The backbone uses (reconfigurable) optical add and drop modules, (R)OADM, to add or drop different signals to/from the access networks. In order to drive the access network and translate between its protocol and the one running in the core, an optical line terminator (OLT) is used. This usually means electro-optical conversion, which is disruptive for QKD. Finally, the backbone can be connected to other rings or long-haul networks.}
\label{fig:metro}
\end{figure}

Metropolitan networks aim to cover the area of cities, with a typical span from a few to several tens of kilometers \cite{MAN_02}. A common architecture of a metropolitan optical network (MON) foresees a division into core and access networks, as depicted in Fig.~\ref{fig:metro}. It should be noted that, actually, the design and topologies in a MON could be more elaborated due to, for instance, external constraints, limited resources, or to the growing needs of the carrier company. However, for the sake of clarity we will stick to the network architecture just outlined as a typical one, denoting it as a \textit{canonical} MON.

In MONs, signals are commonly multiplexed using two well-known approaches: time-division multiplexing (TDM) and wavelength-division multiplexing (WDM). WDM has the advantage of allowing the simultaneous transmission of signals over a single fiber by using different wavelengths (channels), thus increasing the total communication bandwidth. In this work we focus only on this second approach. In addition, the wavelength will be also used to address different users over a particular path.

The standardized use of WDM defines a grid of channels, each with a central wavelength, uniformly arranged in the optical spectrum. Depending on the spectral distance between adjacent channels, WDM can be coarse WDM (CWDM) or dense WDM (DWDM). CWDM is composed of 18 channels spaced from 1270 to 1610~nm and each occupying 20~nm (O, E, S, C and L bands). DWDM is mainly limited to the 1550~nm region (S, C and L bands) and, depending on the chosen grid, channel separation ranges from 100~GHz (or multiples) down to 12.5~GHz ($0.8$-$0.1$~nm) to accommodate from 40 up to hundreds of channels \cite{Ohara_06}.

\subsection{Access network}

An access network follows a point-to-multipoint topology to connect many final users to the core using a simple fiber infrastructure of a few tens of kilometers \cite{Ramaswami_09}. They are typically deployed in the so-called fiber-to-the-home (FFTH) architectures with cables containing multiple optical fiber strands and using passive optical technology (i.e. passive optical networks or PON). In a PON, an optical line terminator (OLT), with direct access to the backbone, is connected through a single fiber to a network component (NC) with $N$ outputs, which in turn is connected through a non-shared fiber to $N$ optical network units (ONU) located at the user's premises. The NC is assumed to be close to the ONUs, thus reducing the amount of non-shared fiber used. Depending on the multiplexing technology used, communications between a particular ONU and the OLT, and vice versa, are addressed either using a specific wavelength or time slot (WDM or TDM, respectively) that differentiates it from its neighbors.

In a typical TDM-based access network (e.g. Gigabit-capable PON or GPON), a beam splitter is used as the NC to connect multiple users. This introduces 3~dB of losses each time the number of users is doubled. Hence, a network of 32 users has a minimum of 15~dB losses in the NC. Instead, in a WDM-based approach (e.g. WDM-PON), the splitter is replaced by a wavelength multiplexer. This is typically an arrayed waveguide grating (AWG), which has less insertion losses than the splitter (e.g. a 32-channels AWG has $\approx 3$~dB). Moreover, losses do not grow by much when adding more channels. This allows to increase the number of users while maintaining the same overall loss budget. Another key advantage of the AWG that will be used in the present approach is its cyclic behavior: through each output port, not only a single wavelength can be used, but also its periods in the upper and lower spectrum. Despite not being standardized, the common usage is to take advantage of this characteristic and use two spectrum bands to separate downstream and upstream signals \cite{Park_04}.

\subsection{Core network: Backbone}

Different access networks are connected through a core network or backbone that in a MON is typically a ring. A first-level backbone is composed of $M$ nodes covering all the metropolitan area, where each backbone node is connected to the OLT of one or more access networks. Signals within the ring are wavelength multiplexed and a (reconfigurable) optical add-drop multiplexer, (R)OADM, is used at the backbone node to add and drop different channels, i.e. add or extract wavelengths to/from the ring. The connection between core and access networks typically includes an electro-optical conversion, since the protocols and technologies can be very different. However, when the backbone and the access network are both based on optical technology and WDM, they can also be directly connected in the optical domain, thus opening the possibility to support quantum communications. This allows for a realistic network where QKD emitters can connect to different receivers (even if different QKD protocols are used~\cite{Korzh_13}).

Furthermore, a ROADM can also connect the core to a long-haul network in order to reach distant networks. However, we will not consider here this scenario, since the distance in these settings exceeds the loss budget of actual QKD systems.

\section{Multiplexing QKD systems in a MON}
\label{sec:QKD-MON}

The main thrust underlying the scheme above comes from the need to reduce cost while maximizing network throughput, resiliency and flexibility. In the same spirit, we will use WDM technology as a base to construct a QKD network. The creation and stabilization of a quantum channel is a challenging task that imposes strong requirements on the infrastructure. Quantum channels are easily degraded because of photon absorption or stray photons coming from classical signals in the same fiber. Moreover, technologies that could overcome these problems, like quantum repeaters, are still in their infancy \cite{Briegel_98, Tian_10, Osorio_12}. Hence, the communication must follow a direct optical path, with always the same wavelength and within the loss budget of the QKD system.

The objective of the proposed network is to provide an easy to deploy and maintain infrastructure, supporting many non-interfering quantum channels. By sharing the infrastructure among many users, QKD becomes more price-competitive and increases its potential market share. To this end, the network is designed to use in a shared way the very costly dark fiber that is already deployed and as much commercially available optical equipment as possible. It is to be noted that, in most settings, the cost of hiring or deploying from anew a dark fiber offsets by a long margin the cost of the QKD devices themselves. To limit the interference with classical communications signals, we define in principle the QKD network only for QKD purposes, i.e. at first only quantum and \textit{service} signals will be allowed. By service signals, we mean the classical ones used to keep the QKD devices working (interferometer stabilization, synchronization, etc.). In this first approach, a pair of QKD devices only need a quantum channel and a service channel, both directed from emitter to receiver, in order to establish a quantum link. After studying the restrictions imposed by the service channels in Sec. 4, we will discuss the possibilities of adding further channels such as the ones for the classical key distillation protocols in QKD, for cipher-text transmission, or even for purely classical communications unrelated to the purposes of the QKD equipment.

\subsection{Bands structure and channel plan}

In comparison with classical signals, quantum signals are extremely weak. Even with a QKD system working at a 1~GHz rate, the power difference is $\approx 70$~dB. Therefore, the noise generated by classical signals drastically impedes quantum transmission by reducing their signal-to-noise ratio (SNR). In order to avoid this problem, instead of placing all signals together in the same band, we separate them spectrally as already discussed in previous works \cite{Townsend_97a, Townsend_98, Toliver_04, Runser_05, Chapuran_09}. In particular, we define a \textit{service band} at the S, C and L bands ($\approx$ 1500-1600~nm), and a \textit{quantum band} at the O band (1260-1360~nm). The distance between channels in the same band will depend on the specific ITU grid used for the implementation. It might not seem an optimal choice to move the quantum signals to the O band since fiber losses are slightly bigger ($\approx 0.1$~dB/km more), but actually the main source of losses in a MON comes from components such as splitters, filters, multiplexers, switches, etc., and they are similar across the bands (see Tab.~\ref{tab:losses}). Henceforth, we will use the optical loss as the reference value when comparing different proposals instead of the distance. Beyond having well separated wavelengths for the quantum and classical signals, the motivation behind this choice is the ability to use existing DWDM commercial equipment for the classical service signals, which is backed by a mature industry. For example, a possible implementation of the schema could use standard and readily available small form-factor pluggable transceivers for the classical signals in the DWDM 100~GHz grid in the C band that simply do not exist in the O band.  These would be very expensive to commercially manufacture without the high market demand that drove the development in the C band. At the same time, the manufacturing of QKD equipment can be carried out in the O band as it is in the C band. QKD components such as single-photon detectors~\cite{idQuantique} or adequate lasers for the attenuated single photon sources with similar performance exist in both bands. Of course, the opposite choice: classical signals in the O Band and quantum in the C could be possible and it is just a straightforward modification of the proposed network, but we think that the cost of the equipment would favor the present choice in most designs. On the other hand, since putting the quantum channel in the O band introduces more losses, the measurements in the present paper could be considered, from a secret key-rate performance perspective, as a worst case between the two possible choices.

\begin{figure}[htbp]
\begin{minipage}[b]{0.48\textwidth}
\begin{center}
\subfloat[]{
\includegraphics[width=\textwidth]{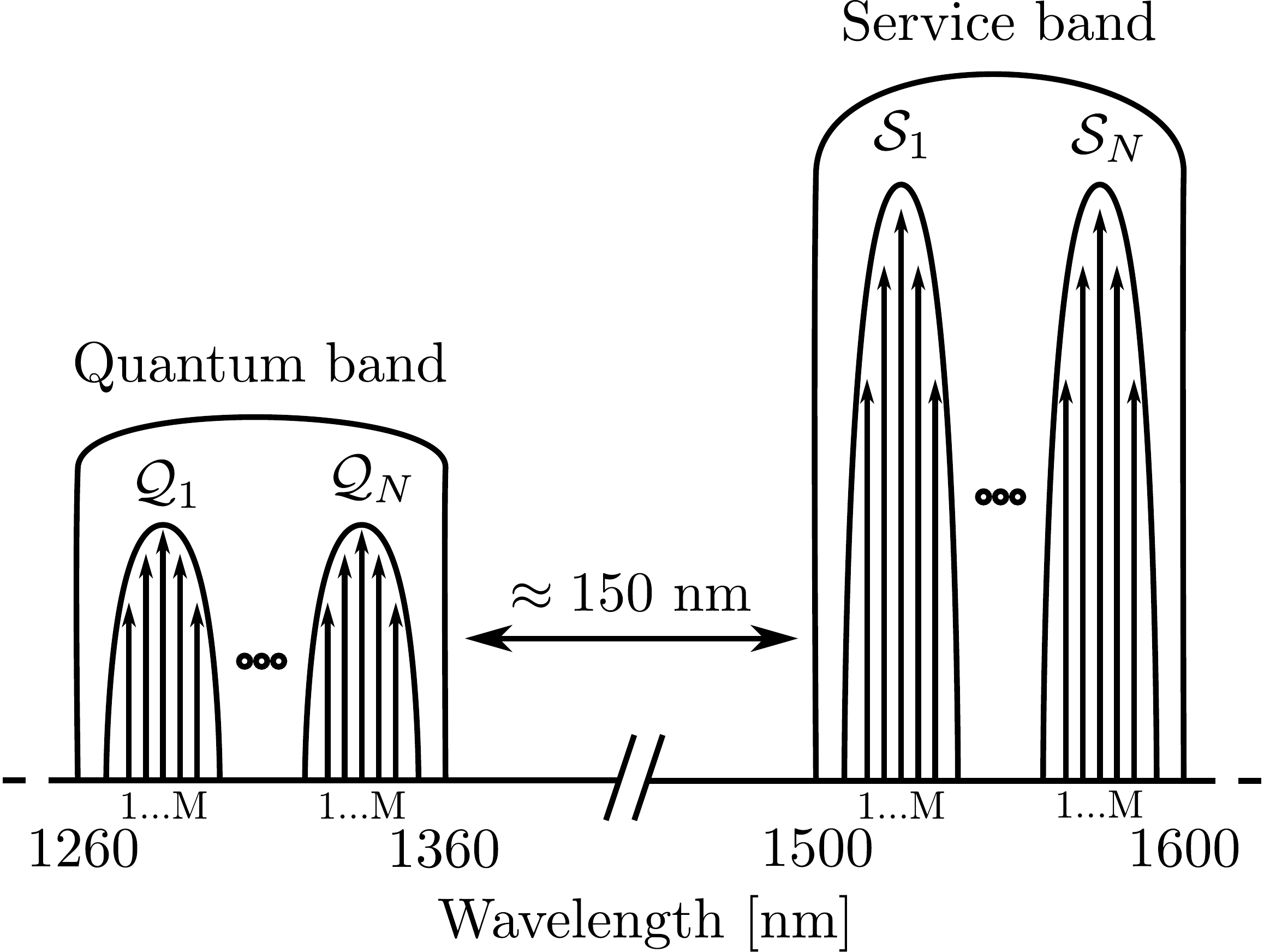}
\label{fig:spectrum}
}
\end{center}
\end{minipage}
\begin{minipage}[b]{0.52\textwidth}
\begin{center}
\subfloat[]{
\includegraphics[width=\textwidth]{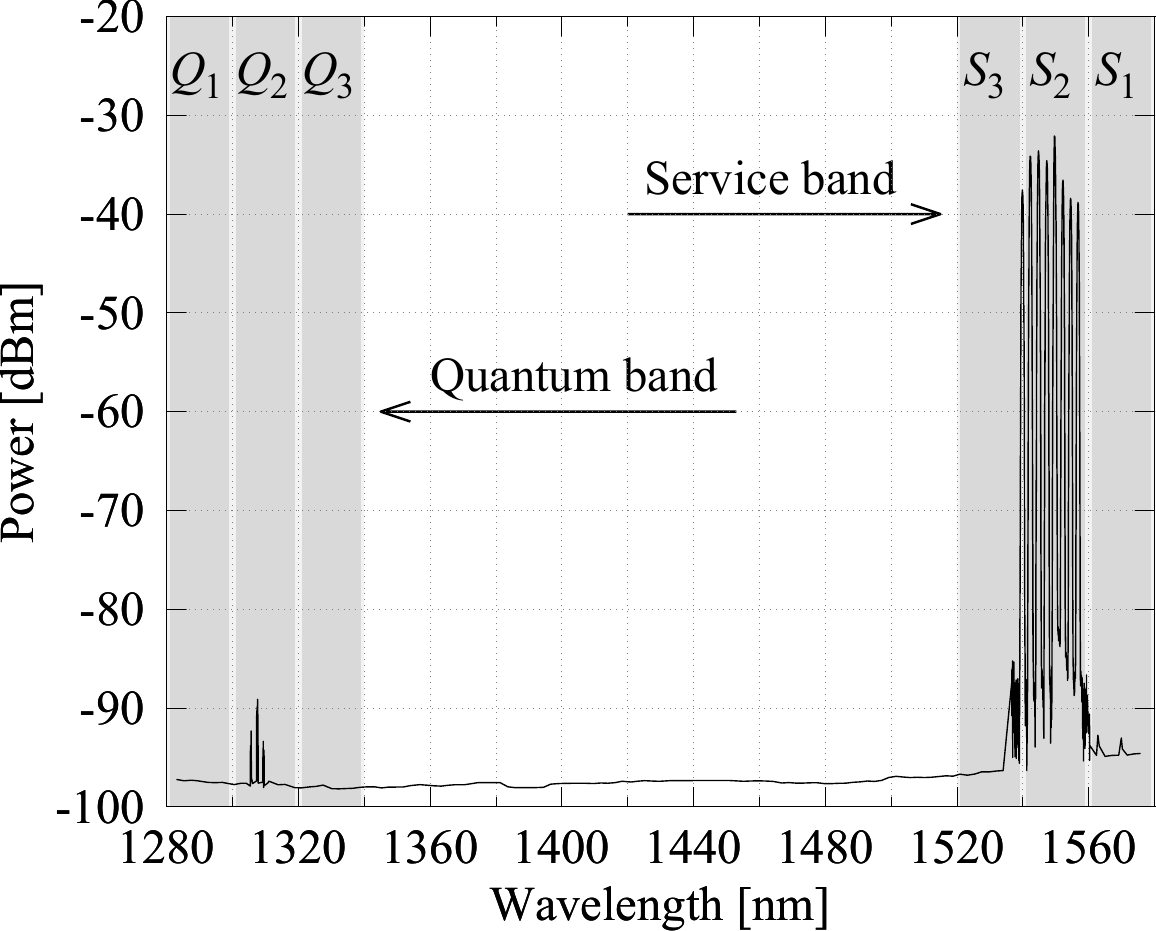}
\label{fig:exp_spectrum}
}
\end{center}
\end{minipage}
\caption{(a) Spectrum of the proposed wavelength-multiplexing scheme. The spectrum is divided in two bands, quantum and service, separated well enough to minimize noise in the quantum band. The first band is located in the O band (13xx) and is used to transport the quantum channel. The second, mainly at the C band (15xx), carries the \textit{service} channels needed to keep the quantum channel and cryptographic protocol working. Each of the two bands is divided in $N$ subbands, named here $Q_{1 \dots N}$ and $S_{1 \dots N}$, for quantum and service, respectively. A pair of quantum and service subbands will correspond to an access network. Each subband carries $M$ channels, represented here as arrows. Channels are chosen in an ITU grid and periods of the AWG. Subbands are selected such that the corresponding wavelengths in the quantum and service band are in the same period, hence both will come out together in the same AWG port. (b) Experimental spectrum of the network prototype. To check the behavior of the network prototype, two signals were fed into the quantum and service subbands Q2 and S2. The subband structure is clearly seen. The different number of channels seen in both bands are due to the input signals used for the test. For a complete description, see Sec.~\ref{sec:prototype}.}
\label{fig:both}
\end{figure}

We will assume a MON where QKD systems are placed at the access networks end-user nodes, as if they were ONUs. In order to distribute the channels among them, we slice the quantum and service band in as many AWG-periodic subbands as there are access networks, and then we assign a pair of quantum and service subbands to each access network. In this way, each QKD device from a QKD system (emitter or receiver) gets a quantum channel and its AWG-periodic service channel. This means that a pair of QKD devices will have available four channels (two in each direction) to run the QKD protocol, although not all are used in our first approach for the service channels. Fig.~\ref{fig:spectrum} shows the schematic spectrum resulting from this approach. Therefore, the selection of a certain pair of wavelengths by a QKD device will also select a specific access network and, within that network, a specific QKD device. This addressing mechanism also allows an easy filtering of unwanted signals at the receivers side.

\begin{figure}[htbp]
\centering
\includegraphics[width=0.9\textwidth]{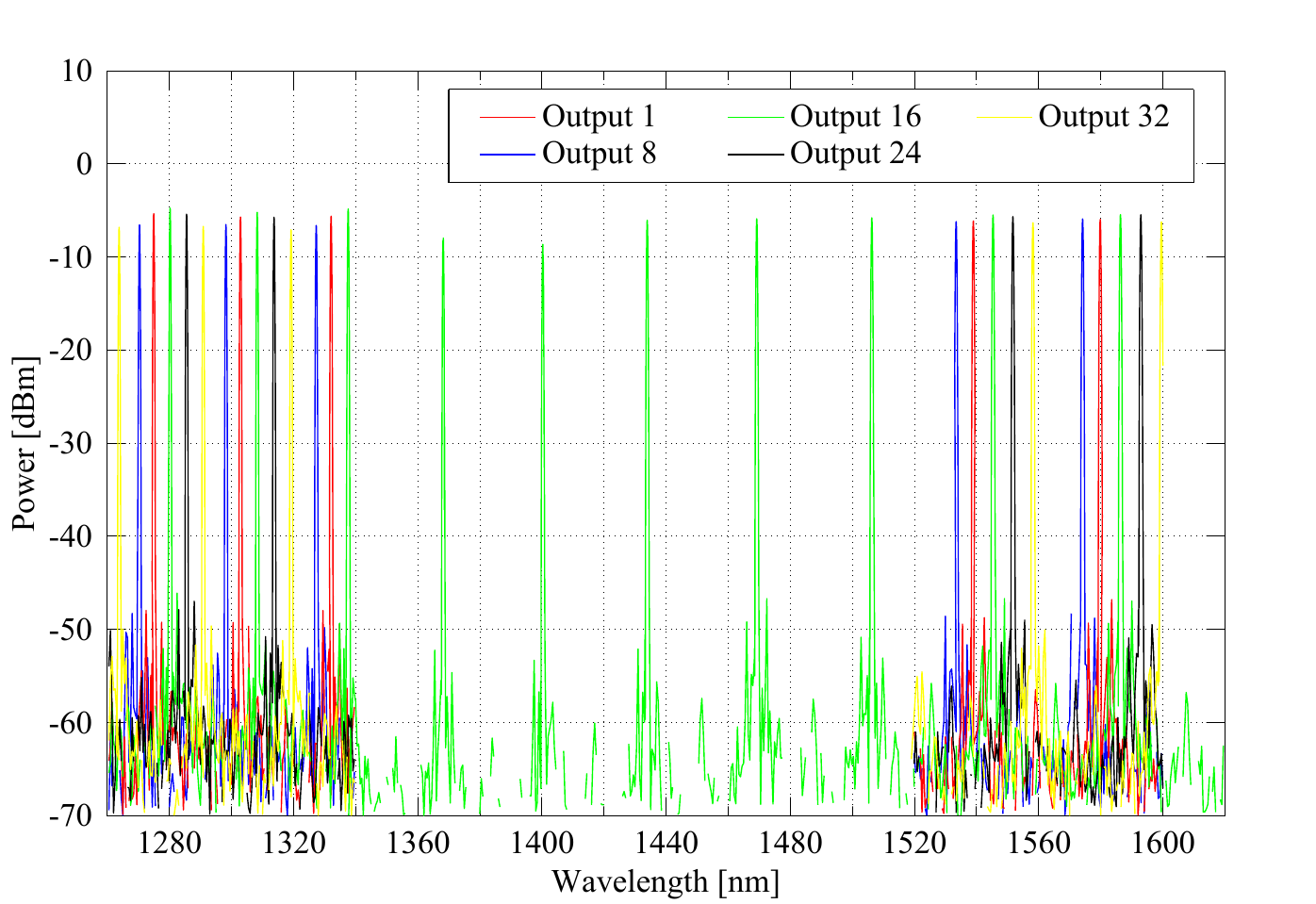}
\caption{Experimental data of the cyclic behavior of a 100~GHz 32-channels AWG, as the one used in the network prototype, in the range 1250-1620~nm. Only outputs 1, 8, 16, 24 and 32 are shown, and only output 16 is presented over the whole range. Channels from the same periodic set have the same color.}
\label{fig:cycles}
\end{figure}

The pairs of quantum and service channels must be routed identically to the same device. To this end, we take advantage of the cyclic behavior of the AWG. For our experimental test bed, we have characterized the periodicity of a standard, telecommunication grade, 100~GHz 32-channels AWG using three tunable lasers to cover the whole 1260-1620~nm range. A given wavelength was fed to the common port and an optical spectrum analyzer was used to measure the output port. In Fig.~\ref{fig:cycles} we present the spectrum obtained summing the outputs 1, 8, 16, 24 and 32. Only output 16 is shown for the full range, including the 1340 to 1520~nm region that separates the quantum from the classical signals and is not used in the proposal. The figure clearly shows the periodicity used to route the corresponding pairs of quantum and service bands to the same destination. We define as a \textit{periodic set} the set of channels that can be used through each output port of the AWG.

\subsection{Simplified network}

\begin{figure}[htbp]
\begin{center}
\includegraphics[width=0.45\textwidth]{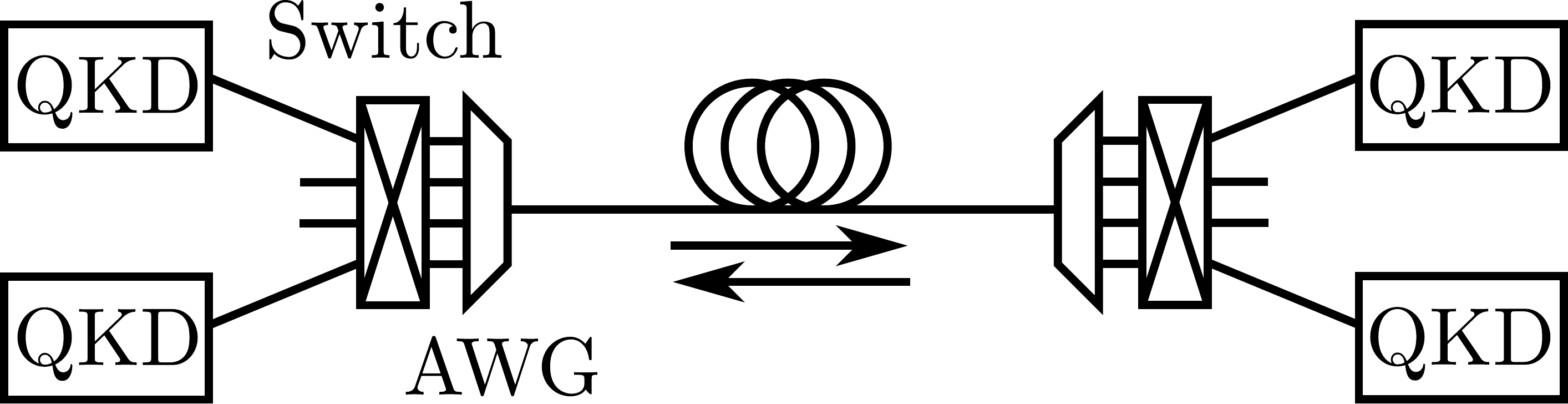}
\caption{Simplified network with two WDM-PON access networks. Only one switch is actually needed to allow a wavelength tunable QKD emitter to use any port of the AWG, and thus communicate in an all to all configuration with any receiver on the other AWG. If a switch is used on each side, as depicted, then emitters and receivers can be freely mixed in both access networks. See text.}
\label{fig:two-access-networks}
\end{center}
\end{figure}

Using this approach, it is straightforward to build a simplified two access networks MON, as depicted in Fig.~\ref{fig:two-access-networks}. In this case, the backbone is just a fiber running from one access network to the other. A wavelength tunable QKD emitter ---one that can use any channel in the quantum band and the corresponding periodic one for the service band--- located behind one of the AWGs could address any QKD receiver located in the other AWG just by changing the pair of wavelengths, and vice versa. The AWG imposes that both QKD devices must be connected to the same output port of their respective AWGs, since ports only allow to pass wavelengths in the same periodic set. This is easily solved by adding an $M \times M$ switch in front of the emitter's AWG. This switch is the only active element in the network. On the other hand, optical switches do not spoil the quantum signal and have very low losses. With this modification, the network is an all-to-all, wavelength addressable and dynamically reconfigurable network, since any QKD emitter can communicate with any receiver at any time by using the appropriate channel and setting the switch accordingly. This network is, however, directional; all emitters have to be located in one access network and all the receivers in the other. If a switch is also added to the receiver's side AWG, this limitation no longer exists and QKD emitters/receivers can be freely mixed and located at any port of any of the two access networks.

\subsection{Backbone nodes and full QKD-MON}

In order to extend this network to a realistic MON, we have to be able to connect more than two access networks. As described in Sec.~\ref{sec:man}, this is done in classical communications using (R)OADMs. In our case, we would need special OADMs able to add and drop simultaneously pairs of bands located in different parts of the spectrum. This has to be done for any subband and in the appropriate periodical sets, introducing the minimal amount of losses as possible and without disrupting the quantum channel. Commercial equipment is not designed to do this, hence we need to devise a dedicated one.

\begin{figure}[htbp]
\begin{center}
\includegraphics[width=0.45\textwidth]{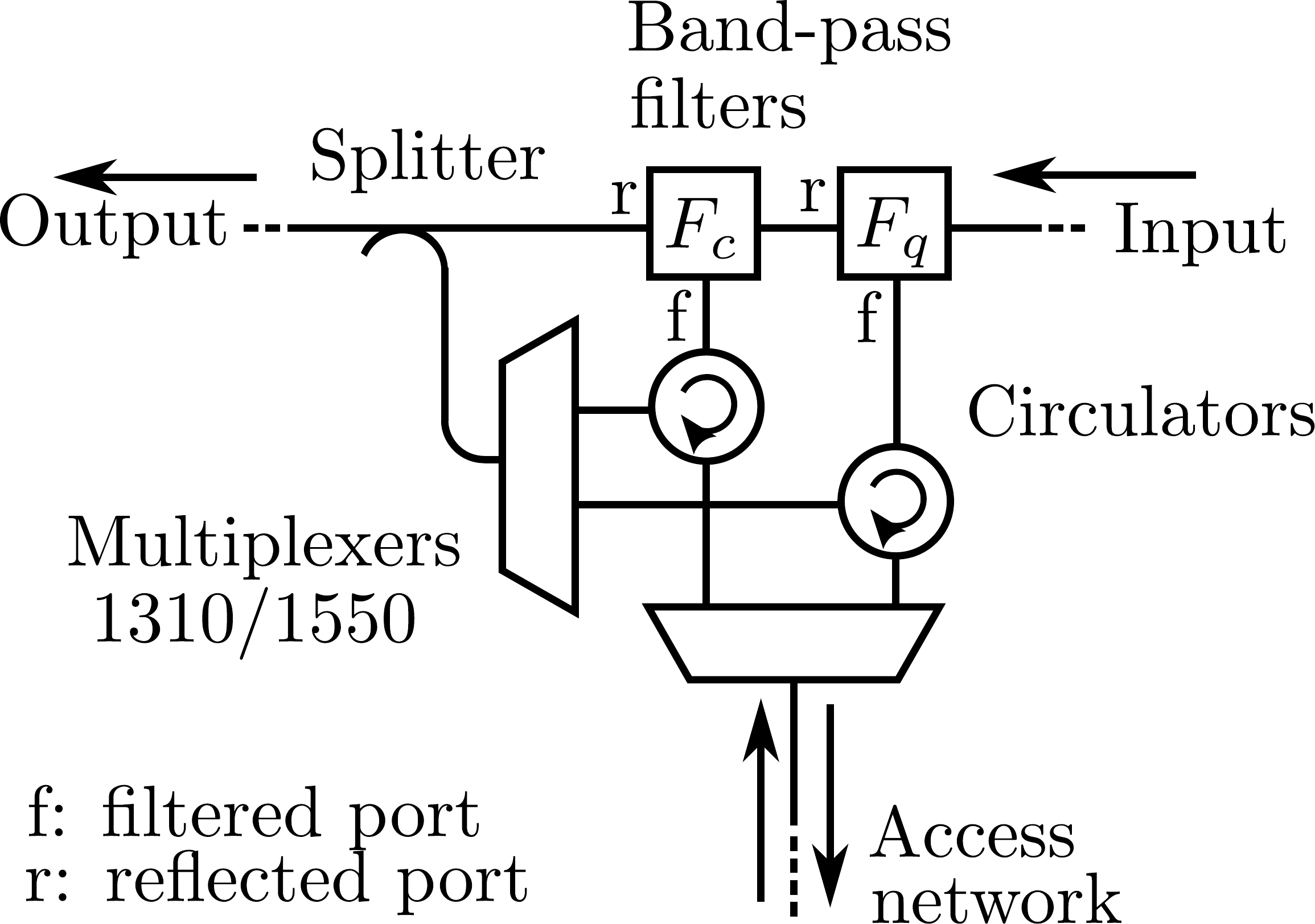}
\caption{Backbone node: OADM designed for the QKD-MON. Built out of common network components, it drops the quantum and service subbands from the ring's signal (input) to the access network, and adds any channel coming from the access network, no matter which subband it belongs to, to the ring (output).}
\label{fig:backbone-node}
\end{center}
\end{figure}

There are several possible designs that can be adapted to different scenarios. Here we adopt a particular one (see Fig.~\ref{fig:backbone-node}) that has low losses and that is easy, reliable and cheap to build. All components used are standard and commercially available, thus able to pass the quality and availability tests required in a real-world deployment. They are also passive. In addition to the already mentioned benefits, the use of passive components in the OADMs has a clear advantage: all paths are always available and no action is required by an external participant to switch between them. As it can be seen in the figure, when the signals enter the OADM, two band-pass filters drop the quantum and service subbands assigned to the access network through the filtered port. These subbands are routed downstream using circulators and they are coupled using a 1310/1550 WDM multiplexer before they reach the AWG in the access network. In the upstream direction, a 1310/1550 WDM mux separates the signal into quantum and service bands. Both are sent, using the same circulators to another 1310/1550 WDM mux that joins them. Finally, they are added to the signals reflected by the band-pass filters using a $1 \times 2$ splitter and injected into the ring no matter which subband they belong to. The key aspect of this OADM is the passband width of the components, since it will determine the specific wavelengths and width of the bands and subbands, hence the addressing and number of channels. Moreover, note that OADMs give a directionality to the backbone network. Hence, the backbone must be a closed ring in order to guarantee communications among all access networks.

The splitter is the component that introduces most losses. These can be reduced by changing the splitting ratio. It can be optimized depending on the number of OADMs that have to be crossed. For instance, for 3-4 backbone nodes, using a splitting ratio of 70:30 reduces the losses about 2 dB in a path crossing the full network, although this is at the expense of increasing the losses in other paths.

\begin{figure}[htbp]
\begin{center}
\includegraphics[width=0.75\textwidth]{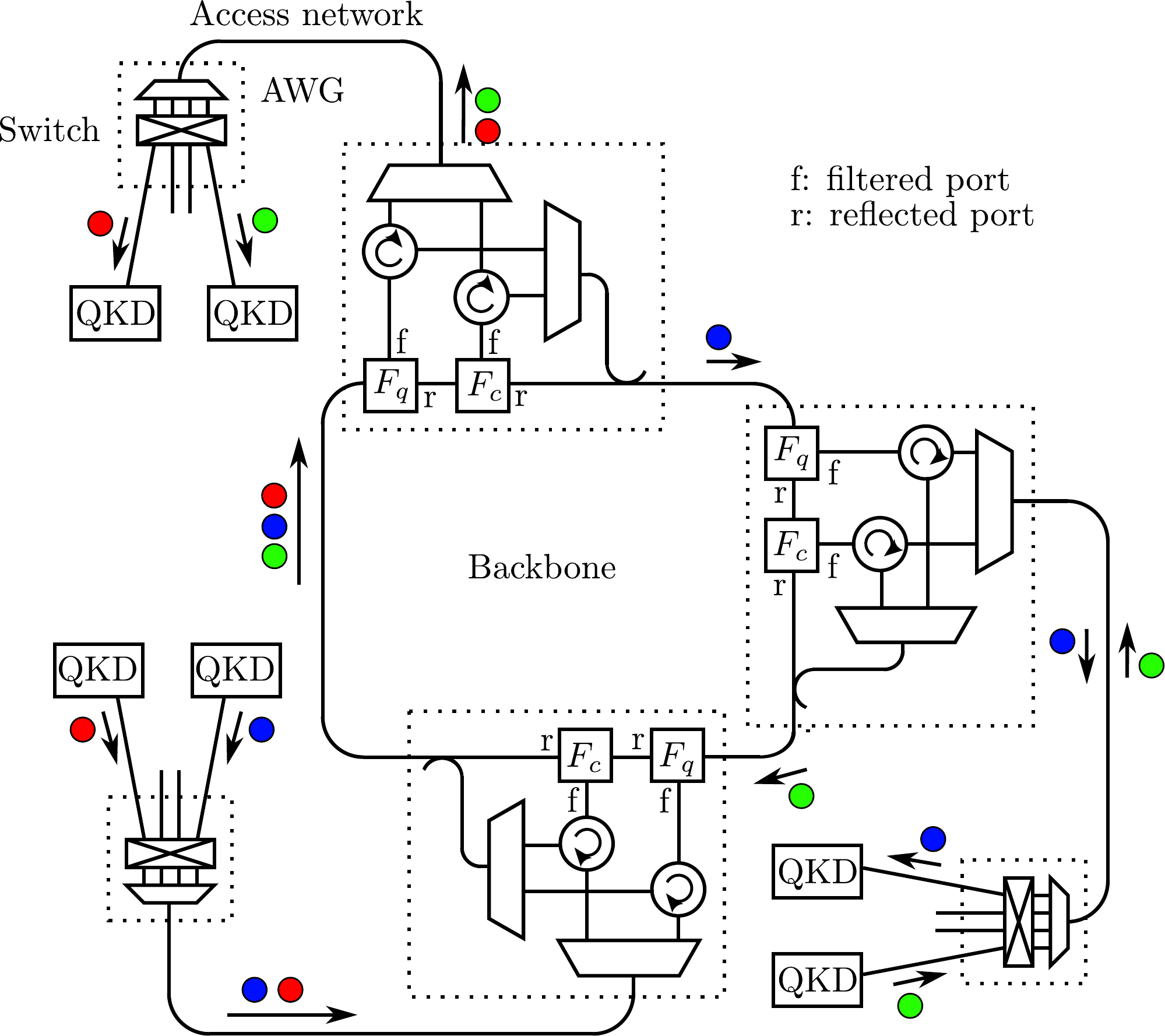}
\caption{Proposed QKD-MON with three access networks. Colored dots are used to illustrate the communication over different paths. Each colored dot represents a pair of wavelengths: one in the quantum band and the corresponding periodic one in the service band that would come out of the same output port of the AWG. The main network components are represented in dashed-line boxes. These devices can be built with out of the shelf commercial components. Note how one access network can communicate simultaneously with the others in a non-blocking way: communications can be performed simultaneously since the network operates employing wavelength multiplexing.}
\label{fig:QKD-MON}
\end{center}
\end{figure}

Fig.~\ref{fig:QKD-MON} shows the result of these modifications: an all-optical QKD-MON based on WDM technology with simultaneous, dynamic, all-to-all communications capability where QKD emitters and receivers are freely mixed in any access network. Colored dots are used to illustrate a pair of wavelengths within a periodic set: one color represents one wavelength in the quantum band and the corresponding in the service band. As an example, it is shown how multiple communications are performed simultaneously. The scheme is non blocking in the sense that QKD devices in different access networks (e.g. the ones at the bottom and at the right) can also address devices in a third access network (top) that is simultaneously being used from the other two networks. OLTs have been removed: they are no longer needed since no conversion of any kind needs to be done and they would disrupt the quantum channel. This means that all upstream or downstream signals go straight to the backbone ring or access network, respectively. Note that there is no short path that links directly two QKD systems in the same access network. There are simple local solutions to this problem. For instance, using a larger switch allows to create return paths (i.e. connect them again to the switch) with the free ports in the side of the AWG.

The losses of the network are shown in Tab.~\ref{tab:qman-losses}. They are calculated using the theoretical values of the components used in their construction. For mere illustrative purposes, we also show examples of full optical paths for scenarios with a different number of OADMs (i.e. access networks) and total fiber length. For instance, a loss budget of approx. 30~dB \cite{Stucki_09, Zhang_10, Dixon_08}, allows a QKD-MON with 3-4 OADMs and a span of 15-20~km. Although there are QKD systems with a loss budget over 40~dB \cite{Takesue_07, Stucki_09b, Wang_12}, we do not consider them practical in real world networks, since they are based on superconducting detectors that need cryogenic temperatures to work. Note that the network scheme remains valid even if QKD technology improves, since a higher loss budget can be directly interpreted as adding new backbone nodes or longer fibers. 

One-way QKD systems \cite{Stucki_09, Zhang_10, Dixon_08, Takesue_07, Stucki_09b, Wang_12} can be used directly in this network, one example could be a system running the coherent-one way (COW) protocol. The most recent implementation of a COW system \cite{Walenta_13} uses a quantum channel (emitter to receiver) together with two classical channels (one in each direction) that carry the service signals and the distillation protocol communication via TDM (in Sec.~\ref{sec:modes}, this advanced approach to the service channel is discussed). Time multiplexing a wavelength in the presented network does not pose any problem and the scheme works flawlessly without modifications. Moreover, the COW system can tolerate delays between quantum and classical signals, such as those originating from the small differences in path that can occur in our OADM node. The only requirements to adapt a COW system to the presented network architecture are: (i) move the quantum channel to the O band, this is feasible by adapting the Faraday mirror and the intensity modulator; and (ii) if addressability is required, use a tunable laser.

Regarding the maximum number of users that the network can serve, it depends on the width of the spectrum bands, subbands and DWDM channel spacing chosen. If CWDM is used for the subbands (passband of $\approx 13$~nm) and a 100 GHz grid, extension of the corresponding ITU DWDM grid, for the channels, the network has a theoretical limit of $4\cdot\lfloor13/0.8\rfloor=64$ users. The first term comes from the maximum number of CWDM channels per band, which is limited by the losses in the O band in the shortest wavelengths and by the need of having the quantum and service, classical signals well separated to avoid interference in the longest wavelengths. Four CWDM channels fit in the wavelength plan without problems. The second term is the passband of the CWDM channel over the DWDM channel spacing in nanometers. Since this value increases for shorter wavelengths (due to the relationship frequency-wavelength), we use the C band as reference ($0.8$~nm). 

The maximum number of users can be increased by choosing a smaller DWDM grid although, in practice, mismatches between the specifications of network components (e.g. CWDM channels and the cycles of the AWG), and the noise in the quantum channel coming from the classical service signals will set the specific limit for a given choice. The noise increases the number of erroneous detections in the single photon detectors and can increase the quantum bit error rate (QBER) to the point of precluding any key exchange. These photons are generated mainly by three physical phenomena: Raman scattering, four-wave mixing (FWM) and crosstalk due to imperfect devices. However, we can eliminate the last two since, due to the separation between quantum and service bands, no signal generated by FWM from the service band will fall within the quantum band \cite{Peters_09}. Likewise, strong service signals that could produce too much crosstalk due to insufficient isolation in the devices can be easily filtered because they are also in other band. The only phenomenon that could spoil the quantum signals is Raman scattering, but then a band separation of approx. 150~nm is enough to attenuate it considerably \cite{Subacius_05, Toliver_04}, as we will see in the next section.

\begin{table}[htb]
\centering
\caption{Losses of typical optical network components. Values are from commercial models available in the market \cite{fiber, flyin, polatis, nortel} that are used for the test-bed in Sec.~\ref{sec:prototype}.}
\label{tab:losses}
\begin{tabular}{l l l}
\hline
Device & Passband & Losses \\ \hline
Single-mode fiber & C band & $0.18$~dB/km \\
Single-mode fiber & O band & $0.32$~dB/km \\
Connectors & --- & $0.2$~dB/pair \\
$1 \times 2$ Splitter & $1270-1350 \& 1510-1590$ nm & $3.6$~dB \\
$1 \times 32$ Splitter & $1270-1350 \& 1510-1590$ nm & $16.5$~dB \\
$4 \times 4$ to $192 \times 192$ Switch & --- & $1$~dB \\
Circulator & $1280-1340$ nm & $0.8$~dB \\
Circulator & $1520-1580$ nm & $0.8$~dB \\
CWDM filter & $\approx 13$ nm & $0.4-0.6$~dB \\
1310/1550 WDM mux & $1260-1360 \& 1460-1560$ nm & $0.5$~dB \\
32-ch AWG DWDM mux & 100 GHz & $3$~dB \\ \hline
\end{tabular}
\end{table}

\begin{table}[htb]
\centering
\caption{Calculated losses for the main network modules of the QKD-MON (according to Tab.~\ref{tab:losses}). Using these theoretical values, we estimate the losses of different full optical paths in terms of OADMs and fiber length.}
\label{tab:qman-losses}
\begin{tabular}{l l l}
\hline
Network component & Losses (quantum) & Losses (service) \\ \hline
32-ch AWG & $3$~dB & $3$~dB\\
Switch & $1$~dB & $1$~dB \\
 OADM (add) & $5.4$~dB & $5.4$~dB\\
 OADM (pass) & $4.8$~dB& $4.8$~dB \\ 
 OADM (drop) & $1.7$~dB & $2.3$~dB \\
10-km path and 2 OADMs & $18.1$~dB & $17.5$~dB \\
15-km path and 3 OADMs & $24.7$~dB & $23.2$~dB \\
20-km path and 4 OADMs & $31.1$~dB & $28.9$~dB \\
30-km path and 5 OADMs & $39.1$~dB & $35.5$~dB \\ \hline
\end{tabular}
\end{table}

\section{Network prototype}
\label{sec:prototype}

\begin{figure}[htbp]
\begin{center}
\includegraphics[width=0.75\textwidth]{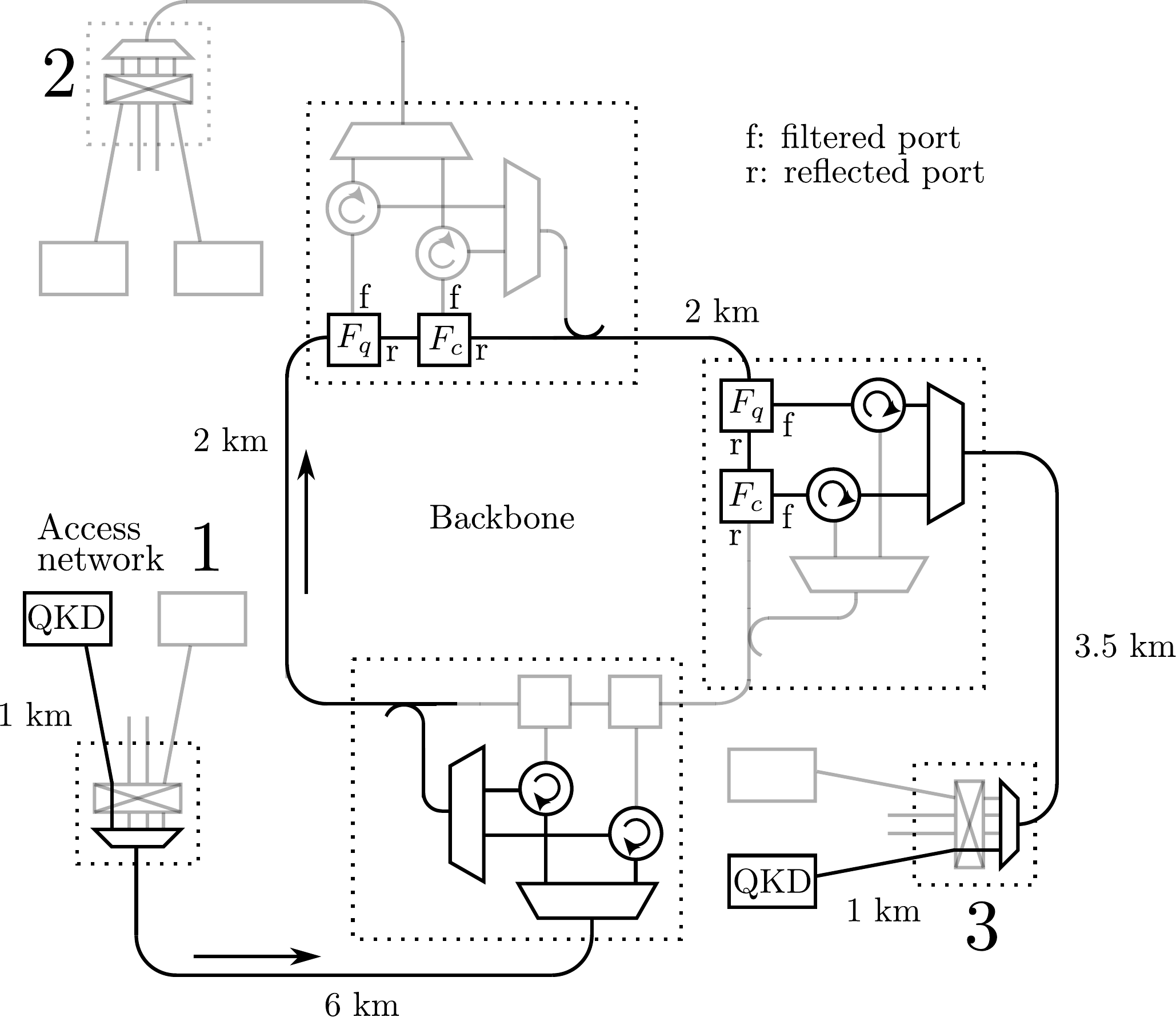}
\caption{QKD-MON test bed: Network prototype with three OADMs built following the design in Fig.~\ref{fig:QKD-MON}. The total length of the fiber is approx. 16~km, a typical span for metro area. A relatively long fiber is used in the access network 1 to generate a high amount of Raman scattering, more than the average for access networks. Overlaid in black is the worst case path in the test bed with respect to losses and generated noise, hence this is the set-up used to perform the measurements.}
\label{fig:scenario}
\end{center}
\end{figure}

The QKD-MON depicted in Fig.~\ref{fig:QKD-MON} has been implemented using the components detailed in Tab.~\ref{tab:losses}. The test bed network is a full-featured quantum metropolitan optical network, including three access networks (labeled from 1 to 3), with static paths. The path used for testing is depicted overlaid on the network scheme in Fig.~\ref{fig:scenario}. It crosses all network components in order to connect access networks 1 and 3. Thus, it corresponds to the worst case scenario in terms of losses and generated noise.

The three access networks are connected by a backbone using single mode fiber with a total length of approx. 16~km. Optical channels are implemented using two spectrum bands (defined by the spectrum bandwidth of the components used at the OADM): 1280-1340~nm for the quantum channels and 1520-1580~nm for the service ones. Two subbands are assigned to each access network: 1280-1300~nm and 1560-1580~nm for the quantum and service channels, respectively, of the access network 1; 1300-1320~nm and 1540-1560~nm to the access network 2; and 1320-1340~nm and 1520-1540~nm to the access network 3. Note that CWDM is used in the backbone for routing subbands, and channels within a subband are redistributed in the access network using a 100~GHz 32-channels DWDM AWG multiplexer. This is a bonus for the practical implementation, since industrial grade components are readily available.

The test bed network has been firstly characterized by measuring the losses. For this purpose, we use lasers emitting at the access network 1 in order to simulate quantum and service signals communicating with the access network 3. We use then an optical spectrum analyzer to measure the peak power of both signals at different points of the network, including the received signals at the end (i.e. the full optical path). The results, given in Tab.~\ref{tab:test-bed-losses}, are consistent with the calculated theoretical values (Tab.~\ref{tab:qman-losses}). In a further working test, the band-subband structure depicted in Fig.~\ref{fig:spectrum} is reproduced. Multiple QKD devices communicating with the access network 2 are simulated using broad-band lasers and an attenuator for the quantum channels. The results are shown in Fig.~\ref{fig:exp_spectrum}. The difference in the number of quantum and service channels is due to the different width of the lasers used for each band.

\begin{table}[htb]
\centering
\caption{Measured losses in the quantum and service band for the OADMs and for the full optical path in the QKD-MON test bed (see Fig.~\ref{fig:scenario}).}
\label{tab:test-bed-losses}
\begin{tabular}{l l l}
\hline
Device & Losses (quantum) & Losses (service) \\ \hline
32-ch AWG & $2.34$~dB & $2.45$~dB \\
OADM (add) & $5.98$~dB & $4.91$~dB \\
OADM (pass) & $5.7$~dB & $5.8$~dB \\
OADM (drop) & $1.83$~dB & $2.24$~dB \\
Full optical path & $23.15$~dB & $20.64$~dB \\ \hline
\end{tabular}
\end{table}

\begin{figure}[htbp]
\centering
\includegraphics[width=0.9\textwidth]{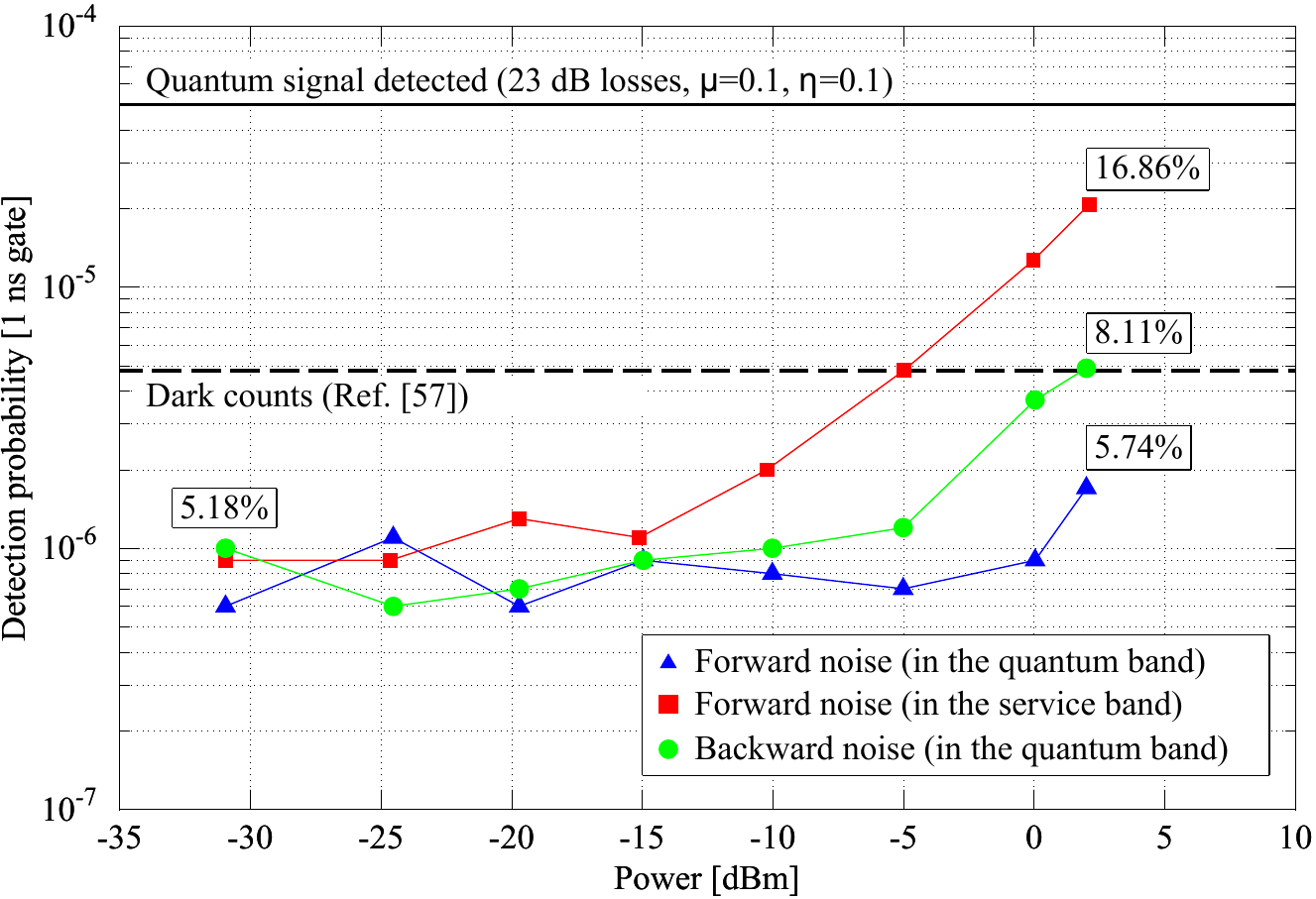}
\caption{Noise measurement done using the setup in Fig.~\ref{fig:scenario}. A laser signal centered at 1520~nm and with power ranging from $-30$ to $+2$~dBm is fed at the access network \#1. The forward noise produced in a quantum channel (1340~nm, triangles) is measured using a SPD at the access network \#3. The backward noise (circles) is also measured by connecting the SPD to the access network \#1. This allows assessing the amount of interference that would reach a QKD receiver coming from an emitter in the same AWG. As a further check, we also measure the forward noise in the service band (1530~nm, squares). To facilitate the comparison, values are normalized considering 1~ns gates. Besides, a quantum signal (mean photon number, $\mu=0.1$, and detector's quantum efficiency, $\eta=0.1$) detected at the access network \#3 and the dark count rate of an SPD \cite{Zhang_10} are also presented. Using these data, a rough estimation of the QBER is shown for multiple points.}
\label{fig:noise}
\end{figure}

Once the network prototype has been checked successfully, we want to find the maximum input power of the service band that does not disrupt the quantum transmission. The critical power is reached when the noise produced by the service signals in a quantum channel together with the intrinsic noise of the single-photon detectors (SPD) used in the QKD yield a QBER equal to the threshold ($11\%$ if we assume that a BB84 with one-way communications is used). In case the power is below the critical value, a QKD link could be established whenever an appropriate QKD system is used, i.e. one able to withstand the 20-30~dB losses\cite{Stucki_09, Zhang_10, Dixon_08}. The power threshold also allows to estimate the maximum number of QKD devices that can operate simultaneously. Using again the full optical path, we have performed several measurements of the forward and backward noise in order to simulate different network configurations (e.g. emitters and receivers mixed together in the same access network). For the forward noise, measurements are carried out at the smallest wavelength separation between quantum and service bands allowed by the channel plan, which is approx. 180~nm (1340 to 1520~nm). This should produce the highest noise levels possible. As a comparison with the schemes where all signals are placed in the same spectral region, the forward noise at the service band is also measured (1530~nm). In both setups, an SPD \cite{idQuantique} is connected to a WDM multiplexer that is connected at the access network 3. The purpose of the WDM multiplexer is to separate the quantum and service bands. At the access network 1, we connect the laser to an erbium doped fiber amplifier in order to try relatively high power configurations (from $-30$ to $+2$~dBm). In this way we can simulate scenarios with a different number of QKD devices. Finally, we measure the backward noise in a quantum channel by moving the SPD and WDM multiplexer also to the access network 1. The measured noise per 1~ns gate in a quantum channel is presented as a function of the input power in the service band for all three scenarios in Fig.~\ref{fig:noise} (note that the intrinsic noise of the SPD has been subtracted). The figure also depicts the noise (dark count rate) of an actual QKD system \cite{Zhang_10} and the expected detection rate of quantum signals. The probability of detecting an emitted single photon is calculated as $1-\exp(-\mu\tau\eta)$, where $\mu$ is the mean photon number, $\tau$ is the transmittance and $\eta$ is the quantum efficiency of the SPD. Therefore, we can estimate the QBER as the ratio of erroneous detections measured with the SPD over the total number of detections. This measurement includes contributions from the dark count rate and the noise generated by the service signals. Calculated values of the QBER of several representative points of the experiment are also shown.

As expected, the forward noise in the service band is higher than in the quantum band. Although the forward noise in the service band is not relevant for QKD operation, it highlights the importance of separating quantum and service bands in the spectrum. The results are consistent with previous findings \cite{Lancho_09, Peters_09, Eraerds_10, Patel_12}, where only a few attenuated classical channels could be transmitted in the same band without disrupting the quantum channel. Also as expected, the backward noise in the quantum band becomes the limiting factor. This is because Raman forward-scattered photons have suffered higher losses (filtering and fiber attenuation integrated over the whole network path). This is something to take into account when a non-directional network is used, since then QKD emitters and receivers can be mixed in the same access network. In this test bed, it is seen that, even with approx. $+2$~dBm power for the service band, an actual QKD system \cite{Zhang_10} can successfully establish a QKD link since the QBER estimation is below the threshold. This overall power for the service band would allow for more than 32~simultaneous service channels of $-13$~dBm without disrupting the quantum channels. For example, in this case, the QBER would increase from $4.37\%$ with no service channel to $5.1\%$ with only one and to approx. $5.74\%$ with all 32 channels being used at the same time. Note that the minimum power in the classical channel has to be chosen carefully to grant a good reception of the service signals. With $-13$~dBm, even in the worst case (highest losses) path, the receiving power of the service channel would be $-34$~dBm. This is strong enough to achieve a data modulation rate of $1.25$~Gbps with a bit error rate no higher than $10^{-9}$ \cite{Patel_12}. Less conservative estimates, using shorter gates at the SPDs (e.g. 100~ps \cite{Zhang_10}), will reduce the noise considerably and thus allow for more service channels or higher data rates.

\subsection*{Modes of Network Operation}
\label{sec:modes}

A data rate of $1.25$~Gbps is obviously wasted if it is used just for service signals which typically have a small duty cycle. It would be highly desirable to go beyond and use the rest of the time for key distillation and/or cyphering. To distil a key, a bidirectional communication is required since classical data has to be sent from the receiver to the emitter. However, note that the backbone ring is directional: a signal originated in the receiver cannot be propagated back to the emitter using the same path. To do this, a signal traveling along the other part of the ring has to be used. Therefore, the receiver has to use a service channel assigned to the emitter. Since, by design, every device in the network has a pair of channels assigned, there is no extra addressing required for these \textit{return channels}; they are already located in the channel plan. However, return channels require a different switch configuration and, thus, they cannot be used simultaneously with the corresponding service channel. This is because, in general, emitter and receiver are connected to different ports of their respective AWGs. Due to the number of signals that need to be generated to produce enough key material to get rid of finite key effects \cite{Scarani_08}, the switching time is not a problem. However, if a simultaneous return channel is necessary, this can be easily taken into account. For instance, in a static version of the network, all channels (i.e. quantum, service and return) can be configured to belong to the same periodic set. If a dynamic addressable network is needed, then the simplest solution is to use different ports of the AWG for each direction. This means that a QKD device will be connected to the switch using two short fibers. This might not be the most economical use of the fiber, however it is not a technical problem since this is a short distance and most installations include spare fibers that could be used for this purpose.

\section{Conclusions}
\label{sec:conclusions}

We have presented a quantum metropolitan network that is, in contrast to existing QKD networks, specifically designed to share infrastructure and use existing optical components in an attempt to make QKD a more cost-competitive technology and lower the barriers to a wider market adoption. We also show that the new modules needed can be built out of  inexpensive, industrial grade and readily available components, without introducing unacceptable losses. The scheme is based on wavelength division multiplexing and addressing, whereby multiple QKD devices are simultaneously connected for transmitting quantum and classical signals. The architecture is a conventional one in metropolitan optical networks, comprising backbone and access networks, although these two segments are directly connected to provide uninterrupted optical paths between all users; a must in order to support a quantum channel. The network allows all to all QKD links and uses standard commercial WDM technology: CWDM for the backbone and DWDM for the access. Except the switches on the user side, needed only if all-to-all dynamic routing is required, the rest of the network is purely passive. This would potentially allow for a cheap, easy and reliable deployment.

The scheme is limited by the loss budget of actual QKD systems ($\approx$ 20-30~dB), but, as discussed above, this is enough for a backbone ring of 20~km with three access networks. This would allow to cover interesting regions in a city and its surroundings. The measurements performed on a prototype network demonstrate that it is capable of supporting at least 32 simultaneous QKD links, each one with a pair of a quantum and a service channel, whereby the latter can support traffic of up to $1.25$~Gbps classical signals. This traffic could include key distillation communication or even cipher text transmission. Classical channels for other purposes could also be included when not all of the possible QKD links are installed. The estimate assumes $1$~ns detector gates: more channels and a higher throughput would be possible if last generation, sub-ns gated detectors are used. We introduced the scheme with discrete, one-way QKD systems but it could, in principle, be extended to entangled pairs and continuous variables, although then the limits could possibly vary. We plan to address in near future the extension of the present scheme to cover all main QKD realizations.

\section*{Acknowledgment}

This work has been partially supported by projects QUITEMAD, \textit{Comunidad Aut\'{o}noma de Madrid}, and Quantum Hybrid Networks, \textit{Ministerio de Econom\'{i}a y Competitividad}, Spain. GAP acknowledges financial support from NCCR-QSIT. AIT acknowledges support by the project QKD-Telco: Practical Quantum Key Distribution over Telecom Infrastructures, within the FIT-IT programme funded by the \textit{Austrian Federal Ministry for Transport, Innovation and Technology} (BMVIT) in coordination with the \textit{Austrian Research Promotion Agency} (FFG). The authors also thank P. Corredera for his assistance regarding the characterization of the AWG, and M. Soto and Telef\'{o}nica I+D for the loan of the OSA, EDFA and fiber prototype network used in this work.

\end{document}